\documentstyle[editedvolume,numreferences]{crckapb}

\newcommand{\stt}{\small\tt}
\newcommand{\dg}{$^\circ$}

\begin{opening}
\title{THE CONTRIBUTION OF MICROLENSING SURVEYS TO THE DISTANCE SCALE}
\author{J.P.Beaulieu \& W.J. de Wit}
\institute{Kapteyn Sterrenkundig Instituut \\
           Postbus 800 \\
           NL-9700 AV Groningen, Netherlands \\
           \stt beaulieu@astro.rug.nl}
\institute{Universiteit Utrecht, Sterrekundig Instituut \\
           Postbus 80000 \\
           NL-3508 TA Utrecht, Netherlands \\
           \stt w.j.m.dewit@astro.uu.nl}
\end{opening}
\runningtitle{Microlensing Surveys and the Distance Scale}
\begin{document}
\begin{abstract}
In the early nineties several teams started large scale systematic
surveys of the Magellanic Clouds and the Galactic Bulge to search for 
microlensing effects. As a by product, these groups have created enormous time-series 
databases of photometric measurements of stars with a temporal sampling
duration and accuracy which are unprecedented. They provide the 
opportunity  to test the accuracy of primary distance indicators, such as Cepheids, 
RRLyrae stars, the detached eclipsing binaries, or the luminosity of the red clump.
We will review the contribution of the microlensing surveys to the understanding
of the physics of the primary distance indicators, recent differential studies
and direct distance determinations to the Magellanic Clouds and the Galactic Bulge.
\end{abstract}
\section{Introduction}
The distance scale problem has been in the center of one of the 
most heated debates in Astronomy this century and, despite
enormous efforts, up to this day large
systematic effects remain between several different distance
indicators. Following Aaronson \& Mould (1986), we recall that  
``the ideal distance indicator: \\ 1. should satisfy small
quantifiable dispersion;\\ 2. should be measurable in enough
galaxies so that it can be calibrated locally, and its intrinsic
dispersion and systematic variation can be tested;\\ 3. should have
a well defined physical basis;\\ 4. should be luminous enough to be
useful at large distance.''
The microlensing surveys give us the opportunity to test the
accuracy of primary distance indicators, such as Cepheids, RRLyrae stars,
the luminosity of the red clump, ... 
Thanks to their systematic  observations of millions of stars,  
it is possible to build complete catalogues of variable stars
or color magnitude diagrams (CMDs hereafter) 
in different galaxies of different metallicities and different 
star formation histories.

Heroic efforts from the ground (Pierce et al. 1994) or
with the Hubble Space Telescope (see other chapters in this volume) give
samples made of a few dozen of Cepheids at maximum in a given very 
distant galaxy with a small number of epochs.
On the contrary, the microlensing surveys provides high accuracy
light curves with extremely good phase coverage for periods of years for
millions of stars (thousands of Cepheids and RRLyrae stars)
in different nearby galaxies.
They provide strong constraints at different metallicities 
for the theory of stellar pulsation and stellar evolution.
Therefore they help our understanding of the physical basis
of these distance indicators. 
They offer a very good basis for differential studies
and for checking the consistency between the different primary
distance indicators.

\section{Microlensing surveys}
Following an original idea proposed with skepticism by Einstein
in 1936 and revived in 1964 by Refsdal, Paczy\`nski (1986)
suggested to probe our Galactic Halo via microlensing effects 
on stars in the Magellanic Clouds. A compact object in the Galactic
Halo passing close enough to the line of sight to a background star 
in the Magellanic Clouds induces an increase in the apparent brightness 
of the star.
This phenomenon occurs owing to the alignment of the observer, the deflector, 
and the background star. If the total mass of our halo is in the form
of compact objects, the probability of a star to be amplified
by at least a factor 1.34 is $0.5~10^{-6}$.
Assuming a co-rotating halo, the time scale 
$\tau_0$ of an event is given by $\tau_0 = 70 \sqrt{{M \over M_\odot} }$ days
where M is the mass of the deflector. Therefore in order to
be sensitive to a wide range of masses for compact objects, 
one should monitor millions of stars and 
be sensitive to events with time scale ranging from hours to years.
The choice of a line of sight towards the LMC and one towards the SMC
would give information about the geometry of the halo (Sackett \& Gould, 1993).
Moreover it is also interesting to monitor stars towards
the galactic center. It provides a proof-of-principle
for the experiment because microlensing is expected from the known
faint end of the stellar luminosity function. Moreover the probability
of seeing MACHOs is not much less than that expected towards the LMC. 
In the early nineties, the technology needed to perform such a search become 
available and three teams decided to enter the game: EROS, MACHO and OGLE.

EROS (Exp\'erience de Recherche d'Objets Sombres -- Aubourg et al. 1993)
is a collaboration between
French astronomers and particle physicists. They adopted two strategies : the first
involves the photographic monitoring of a 25 square degree field in the LMC 
using ESO Schmidt plates. Exposures  have been taken no more than nightly in 
two colors, $B_J$ and $R_C$. About 380 plates have been taken between 1990-1994.
The second approach uses a 0.4m f/10 reflecting telescope and a mosaic of 
16 buttable CCDs covering a field of $1 \times 0.4 $ degree centered 
in the bar of the LMC and in the main core of the SMC. 
Between 1991 and April 1994, about 15 000 images have
been taken in two broad band filters, $B_E$ and $R_E$,
and $250~000$ light curves
with as many as 48 points a night have been obtained.

The MACHO collaboration (Massive Compact Halo Objects -- Alcock et al. 1993) has the dedicated
use of a 1.27m on Mount Stromlo (Australia). They equipped it with a prime focus
reimager-corrector with a dichroic beam splitter which provides a 1\dg \ field in
two passbands simultaneously. In each beam they have one  $2 \times 2$ array
of $2048 \times 2048$ Loral CCDs covering 0.5 square degrees. 
They started the operation in 1992 with the LMC as prime target (monitoring
9 million stars), then the Bulge (monitoring eleven million stars) and the SMC.
They adopted a daily sampling rate, or a few points a week, to be sensitive to
MACHOs in the mass range $ \sim 10^{-3}-10^{-1} M_\odot$. The observations
will stop at the end of this millennium.

OGLE (Optical Gravitational Lensing Experiment -- Udalski et al. 1992)
had on average 75
nights/year a 1m telescope at their availability at Las Campanas
(Chile) between 1992-1995. This telescope was equipped with a $2048 \times
2048$ Ford/Loral CCD chip with a full frame of
$15\times15$ arcminutes. They used two filters,
an I band filter which is closely related to gunn {\it i} and a
V$_{J}$ filter, though the vast majority was done in I. 
Their primary objective was to search for microlensing events
towards the Galactic Bulge. They monitored approximately two million
stars in the Galactic Bulge and reported a total of 19 microlensing
events in their first four seasons.

The first microlensing events towards the LMC and the Galactic Bulge 
were announced almost simultaneously in october 1993 by EROS, MACHO and OGLE.
Now these event are discovered on a nearly daily basis
toward the Bulge and a dozen have been observed toward the LMC and the SMC.

Another microlensing survey started soon after the three pioneers: DUO (Disk
Unseen Objects -- Alard \& Guibert 1997). 
Two hundred Schmidt plates covering a field of 25 square degrees centered
at the galactic coordinates b=-7\dg, l=3\dg \ have been taken in 1994.
Despite the use of photographic plates, this survey has been very successful
thanks to a powerful photometric package developed specifically for the experiment
(Alard 1996).
Part of the EROS schmidt plates have been reprocessed with this package
and a catalogue of
$10 000$ LMC RRLyrae will be released soon (Alard, Beaulieu, Lesquoy \& Hill,
in progress).

EROS and OGLE have been upgraded recently and entered the EROS-2 and the 
OGLE-2 phase.
EROS-2 uses the dedicated Marly 1m telescope at ESO La Silla. The prime
focus is equipped with a focal reducer and a dichroic beam splitter with 
a mosaic of eight CCD $2048 \times 2048$ in each channel. The total field
is $0.7 \times 1.4$\dg. 
The collection of data started in July 1996 with the SMC as prime target.
EROS-2 observes the LMC, SMC, the Galactic Bulge 
(with sampling rate daily, or few points a week).
It performs a search for red dwarves by proper-motion measurements and
a type Ia supernova search.

The second phase of the OGLE project, OGLE-2, uses the new dedicated 1.3 m
Warsaw Telescope at Las Campanas, which saw first light on February
1996, with regular observations starting almost one year later. 
As described in Udalski et al. (1997) the CCD camera used is a
$2048\times2048$ pixels in drift scan mode.
In contrast to the first four pilot years, the OGLE-2 team will be looking for
microlensing not only in our own Galaxy but also towards LMC and SMC.
The data flow is expected to increase 30 fold in comparison with OGLE-1.

At least three other microlensing surveys appeared recently, but have
not provided yet any non-microlensing results to our knowledge.
VATT-Columbia is searching
microlensing in M31 by the image substraction technique (Tomaney 1997). AGAPE
(Ansari et al. 1997) also observed M31, but with a different technique called
pixel lensing  (monitoring of the actual pixel flux). 
MOA (Abe et al., 1997) is a Japan/New-Zealand collaboration planning to search
for microlensing towards the Magellanic Clouds and the Galactic Bulge.
In 1995 appeared two microlensing follow-up networks, PLANET 
(Albrow et al. 1997) and GMAN (closely related to MACHO). 
They are doing accurate, multi-site observations on on going microlensing
events to detect anomalies in the light curves that could be due to
blending, parallax, binaries, or planets.
For details about the results of microlensing surveys, see Paczy\`nski (1996),
Ferlet \& Maillard (1997), and the web sites listed in the bibliographic 
section.

The real strength of microlensing experiments is to realize a systematic 
photometric survey of millions of stars over long period of time.
However we should stress one of their general weaknesses:
these experiments are specifically made to search for
microlensing events. Hence EROS and MACHO adopted wide band filters
in order to get more photons and therefore were able to monitor more stars
improving the statistics for microlensing. In the case of the EROS experiment
for data  taken between 1991-1995, the filters $B_E$, $R_E$ can be transformed
to the standard system (V, I). However, the filters adopted for the EROS-2 survey
 are the result of the convolution of the transmission of the 
dichroic in each path and the CCD response, giving a transmission of 
420-720 nm and 620-920 nm. 
They are so wide that a reliable transformation between this system and a 
standard UBVRI will be very difficult, or even impossible, to determine.
MACHO also have very wide band filter system for reasons similar to EROS
and detailed calibrations have not been published yet.
EROS-2 and MACHO generate homogeneous very large 
databases in their own photometric system,
but comparison with other observations, models, temperature calibrations,
are not trivial at all.

The strategy adopted by OGLE and OGLE-2 is much more attractive than the ones
from MACHO and EROS-2 from a non-microlensing point of view: they
observe with standard filters and provide accurate calibrations in
BVI for their catalogues. Moreover OGLE have made a real effort in
order to make their data accessible as soon as possible. One example 
is the early release of a catalogue of BVI measurements of two
millions stars in the
central part of the SMC as part of the OGLE-2 survey.

\section{A better understanding of the physics of distance indicators}
Thanks to the systematic searching aspect of the microlensing surveys, 
large sample of primary distance indicators have been built in
galaxies of different metallicities. In the case of the pulsating
variable stars, these large samples offer new tests to the theory of
stellar pulsation/stellar evolution and opacity tables and therefore 
will help our understanding of these indicators.

Because the Cepheid period luminosity relation is the corner stone of distance
determination since the beginning of the century, most effort has been
devoted to this class of variable stars. We will also present the
current status of the different studies of RRLyrae variable stars.

\subsection{The Cepheid case}
Cepheids are young, intermediate mass (typically $2-10 M_\odot$),
bright periodic  variable stars. 
These stars have left the main sequence and are in a post core hydrogen-burning phase.
Because of this evolutionary stage,  they lie in an area of the HR diagram, 
the so-called instability strip, where their envelopes are instable to kappa mechanism.
They develope radial pulsations. Their period of pulsation is correlated with 
their luminosity, and this period-luminosity relation (PL) has been used 
as the corner stone in deriving local distances, and extragalactic distances for decades.

A Cepheid envelope is an acoustic cavity in which an infinity of modes of pulsation 
exist. However very few of them contribute to the dynamics of the system; the unstable modes
and the marginally stable modes coupled by resonances to unstable modes.
Resonances are known to play an important role in shaping the light curves. 
Therefore an analysis of the shape of the light curves will give us
some information about the dynamics of these stars and about the resonances between 
pulsational modes.

Two large Cepheid catalogues were built from EROS observations (550 Cepheids in the 
LMC and the SMC -- Beaulieu \& Sasselov 1997 and references therein),
MACHO observations (1466 Cepheids in the LMC -- Welch et al. 1997 and references
therein) and from a pilot campaign of EROS-2 observations ($\sim$900
LMC and SMC Cepheids -- Bauer et al. 1998).

It was known for years that the Cepheids divide into two groups, the  
Classical Cepheids with rather high amplitude, asymmetric curve,
and the so-called s-Cepheids, with low-amplitude symmetric light curves.
Following the suggestion of Antonello et al. (1986), Beaulieu et al. (1995)
showed that this morphological classification is mirrored by a
dichotomy in the period-luminosity plane: the classical Cepheids are fundamental
pulsators whereas s-Cepheids are first overtone pulsators.
Just on the basis of the light curve shape,
it is possible to distinguish between Cepheids pulsating in different modes.

\subsubsection{Classical Cepheids, s-Cepheids, and Beat Cepheids}

In our Galaxy, the LMC, and the SMC, we observed the Hertzsprung progression 
of the changing form of Cepheid light curves due to a 2:1 resonance
 between the fundamental and the second overtone (also known as the bump Cepheids).
This resonance takes place at $10\pm0.5$ days in our Galaxy. 
Using data obtained by the MACHO, EROS, and EROS-2, the resonance takes place 
between 10.5-12 days. The resonance in the 
SMC takes place in the range 10.5-13.5 days (the upper value being poorly 
constrained).

The s-Cepheids (first overtone pulsators) have been observed in the three
galaxies.  They present the same evolution of light curve with period, 
alleged to be the signature of a 2:1 resonance between the first and
the fourth overtone. It takes place at $3.2 \pm 0.2$ days in 
our Galaxy, $2.7 \pm 0.2$ days in the LMC, and $2.2 \pm 0.2 $ days 
in the SMC. One can notice that unlike for fundamental pulsators, 
the resonance is taking place at shorter periods when going to 
lower metallicity. 

So far 73 beat Cepheids have been found in the LMC by MACHO (Welch et al. 1997 
and references therein), eleven in the SMC by EROS (Beaulieu et al. 1997)
and fourteen are known in our Galaxy (Pardo \& Poretti 1997 and references therein).
They are pulsating either in the fundamental and first overtone mode (F/1OT hereafter)
or the  first and second overtone mode (1OT/2OT hereafter).
The SMC 1OT/2OTs are very similar to the LMC ones 
while the SMC F/1OTs have period ratios systematically higher than the LMC 
ones by $\sim 0.01$ which are systematically higher
than the Galactic ones by $\sim 0.01$.

\subsubsection{Searching for consistency between the theory of pulsation,
evolution and opacities}

With two kinds of beat Cepheids, plus the two resonance constraints 
on the classical Cepheids and the s-Cepheids observed at different 
metallicities, we are probing different  depths in the Cepheid envelopes, 
and drawing new strong constraints (similar to helioseismology) 
for the theory of stellar pulsation, stellar evolution and the 
opacity tables at low metallicities. 

When going to lower metallicity, the $\sim$10 days resonance for fundamental
Cepheids occurs at longer periods, whereas the resonance at $\sim$3 days for
overtones occurs at shorter and shorter period.
The period ratios  of F/1OT beat Cepheids increases when decreasing the
metallicity, whereas they are the same for 1OT/2OT for LMC and SMC.

 From a theoretical point of view, one has to keep in mind that when
going to lower metallicities, the evolutionary models will increase
the luminosity at a fixed mass. The increase of mass will lead to a 
diminution of the calculated period ratios. However when going to
lower metallicity, the opacity bump that drives the pulsation will be smaller
and therefore will increase the period ratios for a fixed mass and
luminosity. However since the different modes of pulsations are
probing different depth in the Cepheid envelope, the net effect will
be different from mode to mode. The final result, the observed
position of a resonance center will be a combination of these
antagonistic effects (plus a possible non-linear shift, 
particularly in the case of resonance coupled modes).

The determination of the Cepheids masses has been a long standing 
problem. Serious discrepancies existed between masses from evolutionary 
theory and pulsation theory.
The well-known mass problem of the Cepheids (for a review, see Cox 1980)
led Simon to suggest a revision of the opacities  (Simon 1982;
Andreasen 1988). The use of  improved opacities (OPAL -- Iglesias et al. 1992;
OP -- Seaton et al. 1994)
has substantially decreased the mass discrepancy  but not totally removed it. 
Perhaps more importantly, the bump Cepheids 
have revealed a strong sensitivity to the recent opacities and the mixture.
(Moskalik et al. 1992, Simon \& Kambur 1994).
The strong sensitivity to opacities makes it a useful tool to test the
opacities at different metallicities using extragalactic Cepheids.

The presence of two pulsating modes in the envelope of a Cepheid
gives an anchor for pulsation theory : it is possible to obtain the luminosity 
and the mass  independently of any evolutionary model,
given the temperature and the chemical composition.
Buchler et al. (1996) showed that the implication of the 2:1 resonance
between the fundamental and the second overtone around ten
days for the Bump Cepheids, the alleged resonance around three days 
for the s-Cepheids are difficult to reconcile
with the envelope models at
low metallicity (Z=0.01 and Z=0.004) with the OPAL93 opacities. When going to
lower metallicities the derived masses are too small. 
The discrepancy increases when decreasing the metallicity.
 Whereas a single mass-luminosity relation (ML) is able to reproduce 
the extend of the instability strip, there remains some  
 discrepancy for the Bump Cepheids of the Galaxy, i.e. the mass still differs
by $\sim$10\% as compared to evolutionary models using the same set of OPAL
opacity tables.  This discrepancy (whatever its origin) can be  quantified as 
an increase of the overshoot parameter. However an increase of the
overshoot parameter in the evolutionary models (suggested by Chiosi et al.
1993 as a solution of the Cepheid mass problem) will be in strong disagreement
with other observational constraints.
In contrast for the LMC and the SMC, not only the zero point of the ML
leads to disagreement with evolutionary models mass discrepancy of about $1-2 M_\odot$
but single ML relation cannot render count of the width of the instability
strip.
The OPAL95 version includes several improvements among which the 
incorporation of seven additional chemical elements of the iron group
have been resulted in a further increase of the opacities 
(20\%) compared to the OPAL93 version in the region of the Z-bump which is relevant
for the Cepheids. 
A similar survey done with this new set of opacities
shows that the situation improved, but the discrepancy is not removed.
Meanwhile the results coming from survey of radiative hydromodels at low metallicity
suggest that a strong dissipative mechanism is missing in the envelopes.

Some studies have been focusing on the beat cepheids at different 
metallicities (Morgan \& Welch 1997; Christensen-Dalsgaard \& Petersen 1995;
Antonello et al. 1997; Baraffe et al. 1998).
In these studies 
several mass-luminosities relation from evolutionary calculations
or ad hoc choices are adopted, linear stability analysis of the envelope
with pure radiative, of convection with mixing length theory are performed.
Then they generally concludes that they Beat-Cepheids
period-period ratio planes are reasonably well reproduced for the galaxy, LMC, SMC
metallicities. 
In fact, using their iterative code, Buchler et al. (1996)
showed that at best beat Cepheids give weak constraints on mass luminosity relation.
They provide a useful  set of tests for stellar pulsation theory, but they cannot
be used as a strong constraint on ML relations, unlike the ten-day resonance.

Several attempts to model the resonance at $\sim$3 days for overtone 
pulsators using radiative hydro models (Buchler, private communication;
Antonello \& Aikawa 1995)
have so far failed, showing unphysical ``spikes" in the light curves.

The second overtone mode has been observed in Beat Cepheids. Therefore
it is natural to ask the question whether single mode second overtone 
pulsators exist. No answer has been given yet from an observational 
point of view, but Antonello \& Kanbur (1997) made a survey of hydro models
to study these hypothetical stars and predict resonance positions and characteristics
of the light curves (but again with unphysical features in the light curves).

Over these last years, it has become increasingly clear that there are a number of severe problems 
with radiative models (cf.\ Buchler 1998). A strong dissipative mechanism is missing in the envelope
calculations. The inclusion of a recipe of turbulent convection in stellar envelopes
(Gehmeyr \& Winkel 1992; Bono \& Stellingwerf 1994; Bono \& Marconi 1998;
Yecko et al. 1998) is promising.
The implementation of a relaxation method to obtain non-linear
periodic pulsation and stability analysis of the limit cycles is much
more powerful than very time consuming and sometimes inconclusive
hydrodynamic integrations. The first hydro models of Beat Cepheids
ever computed (Koll\`ath et al. 1998) reproduce period ratio, modal
amplitude and their ratios thanks to full hydrodynamic integration
and the relaxation method. After 30 years of failure
with radiative models, it turns out that the Beat Cepheid
phenomenon is natural and very robust once turbulent convection is implemented. 
Moreover preliminary results show that the Cepheid mass problem is removed
(Beaulieu et al., in preparation) for the SMC and LMC metallicities.
Thanks to the new constraints raised by the microlensing surveys
at different metallicities, significant progress has been made
on the theory of stellar pulsation.
We feel that turbulent convection  puts the nail
in the coffin of Cepheid radiative models...

\subsubsection{A theoretical calibration of the Cepheid PL}

Several efforts are on going in a try to produce a theoretical
calibration of the Cepheid PL relation at different metallicities.
It is definitively a difficult challenge, involving up to date 
opacity tables, proper set of evolutionary models for the different
metallicities, stellar envelope calculations taking into account the
different constraints given by the position of the resonance centers
and the beat Cepheids at different metallicities.
Once {\bf all} these constraints on the modelisation
of the envelope have been met successfully, then stellar atmosphere 
calculations have to be performed and ``put on top'' of the stellar 
pulsation calculations.
Currently  (and for the next few years) these would have to be 
static atmosphere calculations. The complete modelisation of a dynamic 
envelope and atmosphere of a Cepheid is currently still a dream.

The classical study of Chiosi et al. (1993) has been the reference
for distance scale studies over the last years. They computed a large grid 
of Cepheid models varying mass, effective temperature, initial chemical 
composition (Galaxy, LMC, SMC, ...) and mass-luminosity relations 
(with mild or large core  overshoot). 
The linear non-adiabatic stability analysis of the Cepheid  envelopes 
with a treatment of convection by the mixing length theory was performed,
using the (now obsolete) Los Alamos opacities. They derived relations  between 
luminosity, effective temperature and UBVRI magnitudes by using 
theoretical atmosphere models. Among the results from their survey, 
they show that the Cepheid PL show a small dependence of metallicity if 
one uses the V and I bands, whereas the dependence is important
if one uses B and V. This work has been quoted extensively and used as a
strong case to neglect metallicity effects on the Cepheid PL when using 
V and I photometry. However, it is worth mentioning that
nonlinear, nonlocal and time-dependent convective pulsating models are
needed to predict accurate determinations of both blue and red edges
of the Cepheid instability strip and that we should wait for these
models to have a good theoretical understanding of metallicity effects on the PL relations.

Baraffe et al. (1998) present a systematic survey of evolutionary models
and pulsational models in an effort to provide a theoretical calibration 
of the Cepheid PL relation at different metallicities.
They performed an extensive survey of evolutionary
calculations for masses in the range $3-12 M_\odot$ for Z=0.02, 0.01,
0.008, 0.004. The evolutionary calculations are coupled with a
Linear-Non-Adiabatic stability analysis with standard mixing length theory.
They reproduce the period-period ratio diagrams for 
beat Cepheids with good agreement, on the other
hand they do not comments on the position of the resonance centers 
obtained by their modeling. 

Bono \& Marconi (1998) adopted the same ML relation for 
different metallicities, because they consider that the uncertainty
connected with the Helium fraction and the heavy elements in the ML relation,
are of the same order as the decrease/increase in the luminosity 
caused by the metal abundance. Then they compute a survey of hydro models including
non linear non local time dependent turbulent convection, derive 
observable quantities from static stellar atmosphere models, and 
provide a theoretical calibration of the Cepheid PL at different 
metallicities. Even if several uncertainties
remain in the evolutionary calculations, the difference 
of metallicity will imply systematic shifts in the ML relations
which will have a direct impact on any attempt to derive a theoretical 
PL relation. 

\subsection{The RRLyrae case}

Low-mass stars ($<0.8M_{\odot}$) descending the Red Giant
Branch (RGB) and having a rather thin envelope will settle on the
Horizontal Branch (HB) after the onset of He core burning. These stars
are liable to envelope instabilities. These instabilities are driven
by He ionization zone and result in radial pulsations. The
evolutionary loci in the H-R diagram where these stars are found is
termed the instability strip. The instable region in the H-R
diagram depends on the stars exact chemical composition and mass.

These RRLyrae variables which are often found within Globular
Clusters (hence the synonym: cluster variables) are of the low
metallicity population II and have periods ranging from 0.2 to 1.2
days with amplitudes below two magnitudes. Their behavior form an
excellent opportunity for developing and testing current ideas on
pulsation theories, stellar HB evolution and HB morphology.
 
As with the Cepheid case one can distinguish different kinds of
subtypes, depending on the exact mode in which the stars are
pulsating. Phenomenologically one can distinguish these types on the
basis of their lightcurves and periods. Using Fourier decomposition,
the observed population of RRLyrae stars are differentiated into four basic
types. Stars of type RRab are the fundamental mode pulsators, type RRc
constitute the first overtone pulsators, and type RRd are double mode
pulsators. Fourier decomposition is a powerful tool in addressing the
classification method, though good phase coverage and marginal errors
of a lightcurve are required for optimal efficiency of this method.

The internal constitution of the RRLyrae variable stars and the way the
instabilities cause the radial pulsations are reasonably well
understood though understanding of a view long standing problems still
remain. One of these phenomena is the well known elusive variation of
the light curve of RRab's in amplitude and shape, the Blazhko effect.
Different models have been proposed, like Cousens' (1983) of the
oblique magnetic rotator and Moskalik's (1985) of mode resonances, but
none has been appreciated yet. But generally speaking the RRLyrae stars
form a very important distance indicator for Globular Clusters, LMC
and galaxies within the Local Group, due to their regular pulsation
mode and their absolute brightness of approximately +0.8 magnitudes. 
A good calibration for the absolute visual magnitudes with the
pulsation period is of cardinal importance for global distance
indications. 
A synthetic correlation (pulsation equation) of the pulsation period
of the star to its stellar parameters, P$_{0}$=P$_{0}$(L,T$_{eff}$,M),
was reported in the classic paper by Van Albada \& Baker (1971),
marking the start of the controversy between the luminosity of HB
stars derived from evolution theory and from pulsation theory. In view
of this an important discovery was made by Sandage (1982), which is
termed the Sandage-Period-Shift: the increase of the RRLyrae pulsation
period with a decrease of the metallicity.

In the following years extensive studies have been devoted to the
calibration of the RRLyrae absolute luminosity and the parameters it
depends on, observationally as well as computationally. One of the
main objectives is to find the dependency of M$_{v}(RR)$ on [Fe/H]
(Rood 1990; Sandage 1993; Carney, Storm \& Jones 1992) and where
it is shown by Caputo (1997) that this relation is dependent on the
exact morphology of the HB which is quantified by Lee (1989) in the
parameter $(B-R)/(B+V+R)$, where B, R, and V are the number of red,
blue and variable HB stars. She uses synthetic HB computations to
predict the edges of the instability strip and computes the masses
from globular clusters with known [Fe/H] and HB morphology. In this way an estimate
for their distance modulus is made. 

An observational technique for an empirical absolute magnitude
calibration is presented by Kov\'{a}cs and Jurcsik (1997). They
attempt to derive a linear equation for the distance moduli of the
RRab stars on the basis of correlations between Fourier parameters and
$<M_{v}>$, through the basic fact that the lightcurve is in some way
related to the stellar parameters and thus should be reflection of them.

Series of elaborate theoretical investigations have been devoted to
develope theoretical models on the behavior of these kinds of stars
by Bono et al. (1997). With state-of-the-art hydrodynamical codes they present an
atlas of full amplitude theoretical lightcurves accompanied by
predictions for the limits in the H-R diagram of the instability strip
and a updated linear pulsation equation. A study on the
theoretical calibration for RRLyrae with higher metallicity has been
conducted by Bono et al. (1997) as evidence began to stack for high [Fe/H]
popII pulsators (up to solar metallicity). Their results show an
decrease in the amplitude of the first overtone mode with increasing 
[Fe/H] and an opposite correlation for the fundamental mode pulsation.

However the apparent success of hydrodynamical models simulating RRab
stars has been challenged by Kov\'{a}cs \& Kanbur (1998). They show that a
overwhelming majority of the models tested does not follow the empirical
relations derived from observations (e.g. Kov\'{a}cs \& Jurcsik 1997)
regarding the shape of the light curves and the physical parameters. This 
article {\sl ``RRLyrae models  : mission (im)possible''}, shows the actual
limitation of the present theoretical scenarios.

The dawn of large observational photometric databases through the
micro-lensing surveys have shed more light on longstanding problems
within the variable star theories and have helped in constraining
evolutionary and pulsation properties. They brought new developments and
discoveries. One of these new developments concerns the second overtone
pulsating RRLyrae or RRe type. A few stars are suggested to be
candidate RRe (Clement et al. 1979; Walker \& Nemec 1996). 

The MACHO collaboration reports a total number of $\pm$ 8000 field
RRLyrae in the bar of the LMC (Alcock et al. 1996). They argue for a
significant distribution of RRe type stars in the period distribution
of their LMC fields, with a mean period of 0.281 days. They claim that
the lightcurve of this population component shows an asymmetric and
low-amplitude profile, distinguishable from the other type
lightcurves. The MACHO inferences are disputed by Kov\'{a}cs 
(1998). In his paper he makes a case for the RRe's being first overtone
pulsators (RRc's) based on the light curve (no outstanding different
features) and computational evidence from pulsation, evolution and
atmosphere calculations, but he acknowledges the reality of the MACHO
distribution, suggesting that the explanation of the shape lies in the
metallicity dependent HB evolution.

\section{Distance determination from microlensing surveys}

\subsection{Baade-Wesselink distance to the LMC}

Several groups are trying to get distance determination of the
Magellanic Clouds based on different variant of the Baade-Wesselink 
method. The catalogues of Cepheids created by EROS, EROS-2, MACHO 
and OGLE-2 offer or will offer the light curves of a very large number 
a Cepheids that can be used for Baade Wesselink distance determination.
We will just mention two recent contributions to the field.

Gieren et al. (1998) used the near-infrared Barnes-Evans surface brightness
technique with a zero point of the surface brightness color 
relation determined from a large set of interferometrically 
determined angular diameters of cool giants and supergiants
(Fouqu\'e \& Gieren 1997). They are using independently two
magnitude-color combinations (K, J-K) or (V, V-K) and existing radial
velocity curves. Therefore they derive two independent solutions that are
consistent at a remarkable level and got a LMC distance
of $18.46 \pm 0.02$. To include uncertainties for metallicity effects
or other systematics, they give a ``conservative"
LMC distance of $\mu_{LMC}=18.46 \pm 0.06$. 

Krockenberger et al. (1997) developed a new
approach of the Baade-Wesselink method: using HR spectrum and hydro models 
of Cepheid atmospheres they have good understanding of the 
dynamics of the asymmetry of the spectral lines, and therefore can 
provide very accurate radial velocity curves.
They want to reduce the systematic errors in the measurement of the
surface brightness and the temperature by the use of HR spectrum
(instead of color indexes). They then adopt a rigorous statistical 
approach to determine properly the radius and the distance of the target stars. 
 First overtone pulsators from the EROS microlensing survey have 
been observed, and preliminary results presented 
(Krockenberger et al. 1997).
The LMC distance could be determine with an accuracy of 3\%.
In the near future, this method will be used to determine the 
distance of M31 and M33 with an accuracy of 6\% using high
quality photometry obtained by DIRECT (a systematic survey searching 
for Cepheids in M31 and M33 -- Kaluzny et al. 1998; Stanek et al. 1998a)
and Keck spectroscopy.

\subsection{Multi-mode RRLyrae}

Jorgensen \& Petersen (1967) were the first to recognize the
possibilities a double mode pulsating star would open. As they write
in their paper, these stars create the opportunity to make a mass
estimation founded on the ratio of the periods of first overtone to
fundamental mode (by means of the Petersen diagram (PD), which couples
$P_{1}/P_{0}$ to $P_{0}$ on the absiscae). Logically this method yields
important consequences for evolutionary and pulsation scenarios.

Bono et al. (1996) showed that the PD is a valid technique for
estimating the RRd masses. They claim that the best approach for the 
removal of the mass discrepancy existing in the used physical 
route (evolution, pulsation) is by using the non-linear, non-local, 
time-dependent convective models for the RRd variables. (However
they do not compute real double-mode RRLyrae stars. Given the initial
perturbation, they converge either to a fundamental or first overtone
mode).

Observationally with the report of 73 double mode RRLyrae stars by Alcock
et al. (1997), their total number was almost doubled. In the MACHO LMC
fields, RRd's were discovered with fundamental periods between
0.46-0.55 days and $0.742<P_{1}/P_{0}<0.748$, founded on rough
selection criteria. In the studie of these objects they use PD
estimated masses, the theoretical pulsation equation of Bono et al.
(1997) and the assumption of a similarity between the temperature of
RRLyrae stars at the blue edge (Sandage 1993a\&b) of the
instability strip and the temperature of the RRd's, to arrive at a PL
relation.
Finally a distance measurement of the LMC is
straightforward, setting the multimode pulsating RRd based
$\mu_{LMC}$ to $18.48 \pm 0.19$. We recall that the LMC distance
based on single mode RRLyrae (cf.\ Layden, this volume
and references therein) is $18.28 \pm 0.13$.

\subsection{Discovery of an extension of the Sagittarius dwarf Galaxy}

Alard (1996) analyzed the Schmidt plates data obtained for the DUO   
microlensing survey. They cover a field of 25 square degrees centered at the
galactic coordinates b=-7\dg, l=3\dg. He discovered 1466 RRLyrae displaying
a bimodal distribution of magnitude with two clumps separated by 2.3 mag.
He assumes that all RRab have the same  color at minimum light and correct
for extinction. It even reinforce the bimodality of the distribution.
Moreover stars in the two peaks of the distribution follow different
period histograms, indicating a different metallicity.
Most of the RRab belong to the Bulge and
313 stars are concentrated at 24 kpc, which is
consistent with an extension of the recently
discovered Sagittarius dwarf galaxy (Ibata et al. 1994).
Mateo et al. (1995) measured the distance of this galaxy with CMDs to be
$25 \pm 2.8$ kpc. Mateo et al. (1996) and Alcock et al. (1997c) found that
the Sagittarius dwarf galaxy has an elongated main body extending far
from his core, for more than 10 kpc.

\subsection{The red-clump method}

The red-clump stars are the counterpart of the older horizontal branch
in globular clusters and represent a post Helium flash stage of stellar 
evolution (Chiosi et al. 1992).
Paczynski \& Stanek (1998, PZ98) proposed to use the luminosity of
the red clump as a distance indicator. 
They compared the absolute magnitude in 
the I band of about 600 nearby red-clump stars observed by Hipparcos with
accurate trigonometric parallaxes in the solar neighboroud and apparent
magnitude of red-clump stars observed by OGLE in the Baade window to
have a single step determination of the galactocentric distance.
Empirically they found that the average I band magnitude
of clump stars does not depend on their intrinsic color in the range
$0.8 <(V-I)_0 < 1.4$ in the Baade Window. Then they assume no reddening
for their calibrator (clump stars in the solar neighborhood) and
use the reddening maps of Stanek et al. (1996) and Alcock et al. (1998a)
for the Baade window. We stress that  one of their key assumption
is that the two populations of the red clump (calibrator and target) 
follow the same luminosity function. Then they directly compare the
two populations. They get a distance determination to 
the galactic center of $R_)=7.97 \pm 0.08$kpc.

The OGLE-2 team observed four drift scan strips in the SMC, each of them
covering $14.2' \times 57'$ (100 000-150 000 stars per field) and 
4  drift scan strips in the LMC (same angular size, about 200 000 stars 
per strip) in BVI. They built the color magnitude diagrams for these fields.
They adopt a mean reddening of E(B-V)=0.09 for their  SMC fields and
used the reddening maps from Harris et al. (1997)
for their LMC observations. They use the red-clump method following the
precepts of PS97 and derive a very short distance to the LMC and to the SMC,
about 0.4 mag shorter than the generally accepted distances.
$\mu_{LMC} = 18.08 \pm 0.03 \pm 0.12$ and 
$\mu_{SMC} = 18.56 \pm 0.03 \pm 0.06$.

Stanek et al. (1998b), using an independent data set
of LMC observations over a wide field of $2 \times 1.5$, applied
exactly the same method and reached a similar conclusion,
$\mu_{LMC} = 18.065 \pm 0.031 \pm 0.09$.

Beaulieu \& Sackett (1998) showed that the red clump observed by Hipparcos
is well reproduced by the isochrones from Bertelli et al. (1994) with
the distance derived by PZ98, but adopted a LMC distance of 18.3 as 
a best match of the LMC red clump.

Cole (1998) and Girardi et al. (1998) proposed a detailed study of
possible systematic errors in the distance determination using the red-clump
method to show that, with the current evolutionary calculations,
it is not reasonable to assume that the luminosity function of the red
clump does not depend on age, chemical composition, mass loss, star 
formation history. Cole proposes some corrective terms to the determination
of Udalski et al. (1998) and derives a LMC distance of $18.36 \pm 0.18$.

\subsection{Detached eclipsing binaries}

Hilditch (1995) and Paczy\`nski (1996) showed that the observation of detached 
eclipsing binaries with deep narrow primary and secondary eclipses, 
without anomalies in the curve, combined with follow-up spectroscopy is a 
very accurate primary distance indicator.
However such systems are rare and difficult to detect.
Grison et al. (1995) present a catalogue of 80 eclipsing binaries discovered 
by EROS in the bar of the LMC, Alcock et al. (1997a) present the MACHO catalogue of
611 eclipsing binaries in the LMC. These two catalogues provide good candidates that 
could be use for an accurate distance determination of the LMC. 
Pritchard et al. (1998) based on the observation of two systems, 
derive a distance to the LMC of  $18.44 \pm 0.07$.
Kaluzny et al. (1995) discovered eclipsing binaries at the main sequence turn-off
point of Omega Centauri. The accurate observations of these stars will not only
provide a distance determination to this cluster, but will also give us strong 
constraints on stellar evolution calculations.
We also recall
that detached eclipsing binaries are being searched in M31 and M33
in the framework of the DIRECT  project (Kaluzny et al. 1998;
Stanek et al. 1998a).

\section{Differential studies}

Homogeneous catalogues of large  number variable stars and CMDs from microlensing surveys 
in different galaxies of different metallicities start to become available. 
They offer the opportunity to test the consistency between different distance
indicators or to realize sophisticated differential studies.

\subsection{EROS : metallicity effect on the Cepheid PL relation}

The same method (Madore \& Freedman 1991),
based on multicolor photometry to determine reddening corrected Cepheid 
distances, has been adopted by HST distance scale
programs (Freedman et al. 1994; Tanvir et al. 1995; Sandage et al. 1994).
It is assumed that the Cepheid PL relation is universal and that the
Cepheids from the calibrating set and from the target galaxy have 
the same colors. The wavelength slopes of a calibrating set of LMC 
Cepheids are calculated. The PL relation of the Cepheids from the target 
galaxy is slided against it to derive apparent distance modulus
in each band. A true distance modulus of $\mu_{LMC}=18.5$mag, a mean reddening of 
$E(B-V)=0.10$ and a Galactic extinction law with $R_V=3.3$ are adopted for the LMC.
Then using the multicolor apparent distance modulus and a Galactic extinction
law, the total mean reddening and the true distance modulus of the target galaxy 
are determined.

However,  presently available theoretical predictions, suggest that the 
slopes of the PL relation are 
independent of metallicity, only the zero-point is affected, and the metallicity 
effect depend o band pass. If one interpretates the color shift due to
 metallicity  as reddening in deriving the true distance modulus of a target 
galaxy with the method describe above, 
then one makes a systematic error of 
$\delta_\mu = -\delta M_V + R_V \delta(B-V)_0$,  
($\delta(B-V)_0$ is the color change  due to metallicity).

Observational studies have been made since the 70s. For example, an intrinsic 
color shift between LMC and SMC Cepheids has been 
pointed out clearly (Martin et al. 1979). An empirical search for a metallicity
effect (Freedman \& Madore 1990; Gould 1994) in three fields of M31 with
36 Cepheids and 152 BVRI measurements have led to ambiguous results : 
Freedman \& Madore  claimed that there is no significant 
effect. Gould reanalyzed their data with a better statistical 
treatment taking into account the high degree of correlation between 
the measurements and found an effect. However due to the number of
observations, he was not able to solve for a wavelength dependence metallicity
effect, and various systematics of the data set prevented him from 
deriving the size of this effect. 

Beaulieu et al. (1997) and Sasselov et al. (1997)
used the EROS Cepheids data from LMC and SMC to perform an empirical 
test for metallicity effect.
They have high-quality, excellent phase covered light curves for classical
Cepheids and s-Cepheids. Since they pulsate in different modes, they follow 
different PL relations. In the LMC they keep 51 fundamental pulsators 
and 27 first overtone pulsators, and 264 fundamental pulsators and 141 
first overtone pulsators in the SMC.
Thus they have two unbiased samples of Cepheids that fill densely the period
luminosity color (PLC) space, with known difference
in metallicity $\delta [Fe/H]_{LMC-SMC} = 0.35$.

Their method has been be applied independently to 
classical Cepheids and s-Cepheids.
First they compute wavelength dependent slopes for LMC and SMC Cepheids,
these are the same within the error bars. 
They  search for a metallicity effect that depends upon band pass.
They model the data in the PLC plane taking into account the high degree 
of correlation between the measurements. 
The assumption of their model are constant PL slope with metallicity,
and no depth dispersion with the LMC sample. They adopted an LMC true distance 
modulus of 18.5 mag,
a mean reddening of $E(B-V)=0.10$ and a Galactic extinction law with $R_V=3.3$.
The model then has twelve parameters which are :
a linear fit of the PL relation,
a linear fit of the instability strip, 
the distance difference,
the relative reddening difference
and a metallicity dependence on the zero point of the PL relation.

They applied the technique to the Classical and the s-Cepheids 
independently and obtained exactly the same
results. Metal poor Cepheids are intrinsically bluer, and this intrinsic
color change due to metallicity is considered to be reddening when using the
Madore \& Freedman method to derive distances. 
They determined a corrective term due to the  metallicity 
dependence to the Madore \& Freedman distance determination method.
$\delta \mu = (0.44^{+0.1}_{-0.2})\log({ Z \over Z_{LMC} } )$.
Then they discuss its influence of the Hubble constant determination.

Kochanek (1997) applied a generalization of this technique
to do a simultaneous fit to 17 galaxies. He derives also a
significant metallicity correction to the Madore \& Freedman method 
$\delta \mu = (0.4^{+0.2}_{-0.2}) \delta (O/H)$.
Gieren et al. (1998), from observations of galactic Cepheids,
suggest a metallicity correction of $\delta \mu \approx 0.2 \delta [O/H ]$.
Using HST observations of two fields of M101, the HST Key project on distance
scale concluded that metallicity effects are the dominant source of error 
in their error budget, and derive a metallicity correction of 
$\delta \mu = (0.24^{+0.16}_{-0.16}) \delta  [O/H ]$. To nail down this problem, 
Cepheids in about half of the Key project galaxies will also have NICMOS 
observations.

\subsection{OGLE-2 : RRLyrae and the red-clump method}

We would like to summarize a very recent differential study done
by the OGLE team. Udalski (1998) uses OGLE and OGLE-2 observations of 
RRLyrae and red-clump stars in the Baade Window, the LMC and the SMC.
He uses the RRLyrae calibrations from Gould \& Popowski (1998) based
on statistical parallaxes. He derives a weak metallicity dependence 
of the luminosity of the red clump (smaller than the theoretical predictions
of Cole (1998) and Girardi et al. (1998), and found a good agreement between
RRLyrae and red-clump distances which has independent calibrations.
He derives the following distances from the RRLyrae :
$\mu_{GAL} = 14.53 \pm 0.15$, $\mu_{LMC} = 18.09 \pm 0.16$ and  $\mu_{SMC} = 18.66 \pm 0.16$\\
and from the red-clump stars :
$\mu_{GAL} = 14.53 \pm 0.06$, $\mu_{LMC} = 18.13 \pm 0.07$ and  $\mu_{SMC} = 18.63 \pm 0.07$.
Udalski notes as a conclusion ``it is a bit distressing that at the  end
of the 20th century one of the most important topics in the modern astrophysics,
determination of the distance of the LMC -- the milestone for the extragalactic
distance scale -- is a subject of controversy as much as 15\%".

\section{Conclusion}

Already one conference has been entirely devoted to the by-products 
of microlensing surveys in July 1996 at the Institut d'Astrophysique de
Paris ({\sl ``The Astrophysical Return of Microlensing Surveys"}, Eds.\
R. Ferlet \& JP Maillard). Another conference  
({\sl ``The Impact of Large-Scale Surveys on Pulsating Stars Research"}) will 
be held in Budapest in August 1999 and the microlensing surveys will have 
a preponderant part there.

The first contribution of the microlensing surveys to the distance scale problem is by generating
complete samples of variable stars at different metallicities and
color magnitude diagrams made of millions of stars.
They help our understanding of the physics of the different distance
indicators by giving strong constraints on the theory of stellar pulsation, 
stellar evolution and the opacity calculations. 
Maybe more important is the possibility to realize differential studies
between different galaxies and different distance indicators. The goal 
is of course to check the accuracy of the different distance indicators 
used to obtain consistency between them, and finally to nail down the 
metallicity dependence and other  possible systematic effects that 
poisoned the distance scale debate for decades.

Some very significant progress have been made on the Cepheid front.
New strong constraints on the theory of stellar pulsation have been given, 
and we buried the purely radiative codes.  Consistent calculations of evolutionary 
models,  and up to date hydro codes including turbulent convection with the
relaxation method are very promising.
However complete dynamical atmosphere models to evaluate the proper bolometric
corrections, colors and radial velocities for a better comparison with the observation
are still missing.  
The differential studies of Cepheids between the LMC and the SMC based on EROS
data showed  the importance of metallicity effects on the Cepheid distance scale 
contrary to what was generally accepted. 
With two new methods, based on Cepheids observations, it is possible to determine the 
distance up to M31 and M33 with good precision.   
Large catalogues of Cepheids are already available,
or will be released very soon.

Progress have been made too in the understanding of RRLyrae variables.
The RRLyrae have shown to be a very powerful tracer of the population II
in studying the structure of the Bulge,
probing the extension of the Sagittarius
dwarf galaxy or the structure of the LMC/SMC.
Large number of double mode RRLyrae have been already discovered in the LMC, 
and provide a powerful test of stellar pulsation theory. An attempt of distance
determination of the LMC based on calibrations of double mode RRLyrae
by models give a long distance to the LMC. 
On the other hand,
single-mode  RRLyrae still give a short distance to the LMC.

A new method based on the red-clump stars has been introduced. It gives
results with a very small statistical error, whereas questions arise about the
universality of the luminosity function and the influence of
metallicity and of star formation history.  
Thanks to the very large number of clump stars available,
it is very promising, but systematics must be carefully studied.   

Large catalogues of variable stars of all kind are being built in a systematic way.
Among them, detached eclipsing binaries will provide a powerful distance determination to the
local group galaxies in a nearby future. 

Under the cloak of the quest for dark matter, 
gold mines for stellar studies at different metallicities have been found.
Mining is just really starting. \\
\vspace{12pt}


{\small
\begin{thebibliography}{} 
\bibitem[]{} web site EROS : http://www.lal.in2p3.fr/EROS/  
\bibitem[]{} web site MACHO : http://wwwMACHO.mcmaster.ca/  
\bibitem[]{} web site MOA : http://www.phys.vuw.ac.nz/dept/projects/moa/ 
\bibitem[]{} web site OGLE : http://www.astrouw.edu.pl  
\bibitem[]{} web site PLANET : http://www.astro.rug.nl/$\sim$planet/  
\bibitem[]{} Aaronson M., Mould J., 1986,  {\sl Astrophys. J.} {\bf 303}, pp. 1-9.
\bibitem[]{} Abe F., et al., 1997, in {\sl Astrophysical returns of microlensing surveys}, ed R. Ferlet, J.P. Maillard, \'editions fronti\`eres, pp. 75-79.
\bibitem[]{} Alard C., 1996, {\sl Astrophys. J.} {\bf 458}, pp. L17-L20.
\bibitem[]{} Alard C., 1996, {\sl PhD thesis Univ. Paris VI} 
\bibitem[]{} Alard C.  \& Guibert J., 1997, {\sl Astr. Astrophys. } {\bf 326}, pp. 1-12.
\bibitem[]{} Albrow M., et al., 1997, in {\sl Astrophysical returns of microlensing surveys},  ed R. Ferlet, J.P. Maillard, \'editions fronti\`eres, pp. 135-140.
\bibitem[]{} Alcock C., et al., 1993, {\sl Nature} 365, pp. 621-623.
\bibitem[]{} Alcock C., et al., 1995, {\sl Astron. J.} {\bf 109}, pp. 1653-1662.
\bibitem[]{} Alcock C., et al., 1997a, {\sl Astron. J.} {\bf 114},  pp. 3260-3275.
\bibitem[]{} Alcock C., et al., 1997b,  {\sl Astrophys. J.} {\bf 482}, pp. 89-97.
\bibitem[]{} Alcock C.., et al., 1997c, {\sl Astrophys. J.} {\bf 474}, pp. 217-222.
\bibitem[]{} Alcock C., et al., 1998a, {\sl Astrophys. J.} {\bf 492}, pp. 190-199.
\bibitem[]{} Alcock C., et al., 1998b, {\sl Astrophys. J.} {\bf 494}, pp. 396-399.
\bibitem[]{} Alexander D.R. \& Ferguson J.W., 1994, {\sl Astrophys. J.} {\bf 437}, pp. 879-890.
\bibitem[]{} Andreasen G.K., 1988, {\sl Astr. Astrophys. } {\bf 201}, pp. 72-79.
\bibitem[]{} Ansari R., et al., 1997, in {\sl Astrophysical returns of microlensing surveys}, ed R. Ferlet, J.P. Maillard, \'editions fronti\`eres, pp. 47-58.
\bibitem[]{} Antonello E. \& Porretti E.,  1986, {\sl Astr. Astrophys. } {\bf 169}, pp. 149-153.
\bibitem[]{} Antonello E. \& Aikawa T., 1995, {\sl Astr. Astrophys. } {\bf 302}, pp. 105-114.
\bibitem[]{} Antonello E. \& Morelli P.L., 1996, {\sl Astr. Astrophys. } {\bf 314}, pp. 541-546.
\bibitem[]{} Antonello E. \& Kanbur S., 1997,  {\sl Mon. Not. R. astr. Soc} {\bf 286}, pp. L33-L36.
\bibitem[]{} Antonello E., et al., 1997,  {\sl Astr. Astrophys. } {\bf 319}, pp. 863-866.
\bibitem[]{} Aubourg E., et al., 1993, {\sl Nature} 365, pp. 623-625.
\bibitem[]{} Baraffe I., et al., 1998,  {\sl Astrophys. J.} {\bf 499}, pp. L205-L208.
\bibitem[]{} Bauer F., et al., 1998, {\sl Astr. Astrophys. } in preparation
\bibitem[]{} Beaulieu J.P., et al., 1995, {\sl Astr. Astrophys. } {\bf 303}, pp. 137-155.
\bibitem[]{} Beaulieu J.P., 1995, in {\sl Astrophysical Application of Stellar Pulsation}, ed  Stobie R.S., Whitelock P.A. (eds.) ASP Conf. Ser. {\bf 83}, pp. 260-270.
\bibitem[]{} Beaulieu J.P., et al., 1997a, {\sl Astr. Astrophys. } {\bf 318}, pp. L47-L50.
\bibitem[]{} Beaulieu J.P., et al., 1997, {\sl Astr. Astrophys. } {\bf 321}, pp. L5-L8.
\bibitem[]{} Beaulieu J.P. \& Sasselov D., 1997, in {\sl Astrophysical returns of microlensing surveys},  ed R. Ferlet, J.P. Maillard, \'editions fronti\`eres, pp. 193-204.
\bibitem[]{} Beaulieu J.P., Sackett P.D., 1998, {\sl Astron. J.} in press.
\bibitem[]{} Bono G. \& Stellingwerf R., 1994, {\sl Astrophys. J. Suppl. Ser.}, {\bf 437}, pp. 879-890.
\bibitem[]{} Bono G., et al., 1996, {\sl Astrophys. J. } {\bf 471}, pp. L33-L36.
\bibitem[]{} Bono G., et al., 1997, {\sl Astr. Astrophys. Suppl. Ser.} {\bf 121}, pp. 327-341.
\bibitem[]{} Bono G. \& Marconi M., 1998 in {\sl Half a century of stellar pulsation interpretation. A tribute to Arthur N. Cox}, ed P.A. Bradley \& J.A. Guzik,{\sl ASP Conf. Ser.} {\bf 134}, pp. 305-314.   
\bibitem[]{} Buchler R., et al., 1996, {\sl Astrophys. J.} {\bf 462}, pp. L83-L86.
\bibitem[]{} Buchler R.., 1998 in {\sl Half a century of stellar pulsation interpretation. A tribute to Arthur N. Cox}, ed P.A. Bradley \& J.A. Guzik,{\sl ASP Conf. Ser.} {\bf 134}, pp. 220-230.   
\bibitem[]{} Carney B.W., Strom J., Jones R.V., 1992, {\sl Astrophys. J.} {\bf 386}, pp. 663-684. 
\bibitem[]{} Caputo, F.,1997, {\sl Mon. Not. R. astr. Soc} {\bf 284}, pp. 994-1000. 
\bibitem[]{} Chiosi C., et al., 1992, {\sl  An. Rev. Astr. Astrophys. } {\bf 30} pp. 235-285.
\bibitem[]{} Chiosi C., et al., 1993, {\sl Astrophys. J. Suppl. Ser.}, {\bf 86}, pp. 541-598.
\bibitem[]{} Christensen-Dalsgaard J. \& Petersen J.O., 1995,  {\sl Astr. Astrophys. } {\bf 308}, pp. L661-L664.
\bibitem[]{} Clement, C.C., et al., E.E., 1979, {\sl Astron. J.} {\bf 84}, pp. 217-230. 
\bibitem[]{} Cole A.A., 1998 {\sl Astrophys. J.} {\bf 500}, pp. 137-140.
\bibitem[]{} Cox A.N., 1980 {\sl  An. Rev. Astr. Astrophys. } {\bf 18} pp. 15-41. 
\bibitem[]{} Cousens, A., 1983, {\sl Mon. Not. R. astr. Soc} {\bf 203}, pp. 1171-1182. 
\bibitem[]{} Einstein A., 1936 {\sl Science} {\bf 84}, pp. 506-506.
\bibitem[]{} Ferlet R. \& Maillard J.P., 1997 {\sl Astrophysical returns of microlensing surveys}, ed R Ferlet, JP Maillard, \'editions fronti\`eres.
\bibitem[]{} Fouqu\'e P. \& Gieren W.P., 1997,  {\sl Astr. Astrophys. } {\bf 320}, pp. 799-810.
\bibitem[]{} Freedman W. \& Madore B.F., 1990, {\sl Astrophys. J.}, {\bf 365}, pp. 186-194.
\bibitem[]{} Freedman W. \& Madore B.F., 1991, {\sl Publ. Astron. Soc. Pacific} {\bf 103}, pp. 933-957. 
\bibitem[]{} Freedman W., et al., 1994, {\sl Nature} {\bf 371}, pp. 757-762.
\bibitem[]{} Freedman W., et al. 1996, {\sl STScI Colloqium}, ed. M. Livio
\bibitem[]{} Gehmeyr M. \& Winkler K.A., 1992a,   {\sl Astr. Astrophys. } {\bf 253}, pp. 92-100. 
\bibitem[]{} Gehmeyr M. \& Winkler K.A., 1992b,   {\sl Astr. Astrophys. } {\bf 253}, pp. 101-112. 
\bibitem[]{} Gieren W.P., et al., 1998, {\sl Astrophys. J. } {\bf 471}, pp. L33-L36.
\bibitem[]{} Girardi L., et al., 1998, {\sl Mon. Not. R. astr. Soc}  submitted
\bibitem[]{} Gould A., 1994,  {\sl Astrophys. J.}  {\bf 426}, pp. 542-552.
\bibitem[]{} Gould A. \& Popowski P., 1998, {\sl Astrophys. J.} submitted
\bibitem[]{} Grison P., et al., 1995, {\sl Astr. Astrophys. Suppl. Ser.} {\bf 109}, pp. 447-469.
\bibitem[]{} Hilditch R.W., 1995, {\sl Binaries in clusters, ASP Conf.. Ser.} ed. G.Milone, JC Mermillod.
\bibitem[]{} Harris J., et al., 1997, {\sl Astron. J.} {\bf 114}, pp. 1933-1944.
\bibitem[]{} Ibata R.A., et al., 1994 {\sl Nature} {\bf 370}, pp. 194-195.
\bibitem[]{} Iglesias C.A., Rogers F.J., Wilson B.G. 1992, {\sl Astrophys. J.} {\bf 397}, pp. 717-728.
\bibitem[]{} Jorgensen H.E. \& Petersen O.J., {\sl Zeischr. Astrophys.} {\bf 67}, pp. 377-387.
\bibitem[]{} Kaluzny J., et al., 1995a {\sl Astr. Astrophys. Suppl. Ser.} {\bf 112}, pp. 407-428.
\bibitem[]{} Kaluzny J., et al., 1995b {\sl Binaries in clusters, ASP Conf.. Ser.} ed. G.Milone, JC Mermillod.
\bibitem[]{} Kaluzny J., et al., 1998 {\sl Astron. J.} {\bf 115}, pp. 1016-1044.
\bibitem[]{} Kambur S.M. \& Simon N.R. 1994, {\sl Astrophys. J.} {\bf 420}, pp. 880-883.
\bibitem[]{} Kennicutt R.C. et al., 1998, {\sl Astrophys. J.} {\bf 498}, pp. 181-194.
\bibitem[]{} Kochanek C.S., 1997 {\sl Astrophys. J.} {\bf 491}, pp. 13-28.
\bibitem[]{} Koll\`ath Z., et al., 1998, {\sl Astrophys. J.} {\bf 502}, pp. 55-58.
\bibitem[]{} Kov\'{a}cs G. \&  Kanbur S.M., 1998,  {\sl Mon. Not. R. astr. Soc} {\bf 295}, pp. 834-846.
\bibitem[]{} Kov\'{a}cs, G. \& Jurcsik, J., 1997, {\sl Astr. Astrophys. } {\bf 322}, pp. 218-228.
\bibitem[]{} Kov\'{a}cs, G., 1998, in {\sl Half a century of stellar pulsation interpretation. A tribute to Arthur N. Cox}, ed P.A. Bradley \& J.A. Guzik,{\sl ASP Conf. Ser.} {\bf 134}, pp. 52-56.
\bibitem[]{} Krockenberger M., et al., 1997, {\sl Astrophys. J.} {\bf 479}, pp. 875-. 
\bibitem[]{} Krockenberger M., et al., 1997, in {\sl Astrophysical returns of microlensing surveys}, ed R Ferlet, JP Maillard, \'editions fronti\`eres, pp. 231-236.
\bibitem[]{} Lee, Y.W., 1989, {\sl PhD Thesis}, Yale University.
\bibitem[]{} Madore B.F. \& Freedman W., 1991, {\sl Publ. Astron. Soc. Pacific} {\bf 103}, pp. 933-957. 
\bibitem[]{} Martin W.L., et al., 1979, {\sl Mon. Not. R. astr. Soc} {\bf} {\bf 188}, pp. 139-157.
\bibitem[]{} Mateo M., et al., 1995, {\sl Astron. J.} {\bf 110}, pp. 1141-1154.
\bibitem[]{} Mateo M., et al., 1996, {\sl Astrophys. J.} {\bf 458}, pp. L13-L16.
\bibitem[]{} Moskalik P., 1985, {\sl Acta Astron.} {\bf 35}, pp. 229-254.
\bibitem[]{} Moskalik P., et al., 1992, {\sl Astrophys. J.} {\bf 385}, pp. 685-693. 
\bibitem[]{} Morgan S. \& Welch D., 1997, {\sl Astron. J.} {\bf 114}, pp. 1183-1189.
\bibitem[]{} Paczy\'nski B., 1986, {\sl Astrophys. J.} {\bf 301}, pp. 503-506.
\bibitem[]{} Paczy\'nski B., 1996, {\sl An. Rev. Astr. Astrophys. } {\bf 34}, pp. 419-459.
\bibitem[]{} Paczy\'nski B. \& Stanek K.Z., 1998, {\sl Astrophys. J.} {\bf 494}, pp. L219-L222.
\bibitem[]{} Pierce M.J., et al., 1994, {\sl Nature} {\bf 371}, pp. 385-387.
\bibitem[]{} Porretti E., 1994, {\sl Astr. Astrophys. } {\bf 285}, pp. 524-528.
\bibitem[]{} Pritchard J.D., et al. 1998 {\sl Mon. Not. R. astr. Soc} submitted
\bibitem[]{} Refsdal  S., 1964, {\sl Mon. Not. R. astr. Soc} {\bf} {\bf 128}, pp. 295-300.
\bibitem[]{} Rood R., 1990, In {\sl Confontation between Stellar Pulsation and
              Evolution, ASP Conf. Ser. 11}, ed. C. Cacciari and
              G. Clemetini, pp. 11-21.
\bibitem[]{} Sackett P.D. \& Gould A., 1993, {\sl Astrophys. J.} {\bf 419}, pp. 648-657.
\bibitem[]{} Sandage A., 1982, {\sl Astrophys. J.} {\bf 252}, pp. 575-581.
\bibitem[]{} Sandage A., 1993a, {\sl Astron. J.} {\bf 106}, pp. 687-702.
\bibitem[]{} Sandage A., 1993b, {\sl Astron. J.} {\bf 106}, pp. 703-718.
\bibitem[]{} Sandage A., et al., 1994, {\sl Astrophys. J.} {\bf 423}, pp. L13-L16.
\bibitem[]{} Sasselov et al., 1997, {\sl Astr. Astrophys. } {\bf 324} pp. 471-483.
\bibitem[]{} Seaton M.J., et al., 1994, {\sl Mon. Not. R. astr. Soc}  266, pp. 805-828.
\bibitem[]{} Simon N.R., 1982, {\sl Astrophys. J.} {\bf 260}, pp. L87-L90.
\bibitem[]{} Simon N.R. \& Lee A.S. 1981, {\sl Astrophys. J.} {\bf 248}, pp. 291-297.
\bibitem[]{} Simon N.R. \& Kanbur S. 1994, {\sl Astrophys. J.} {\bf 429}, pp. 772-780.
\bibitem[]{} Schwarzenberg-Czerny, 1989,  {\sl Mon. Not. R. astr. Soc} {\bf 241}, 153-165. 
\bibitem[]{} Stanek K.Z., 1996, {\sl Astrophys. J.} {\bf 460}, pp. L37-L41.
\bibitem[]{} Stanek K.Z., et al., 1998a, {\sl Astron. J.} {\bf 115}, pp. 1894-1915.
\bibitem[]{} Stanek K.Z., et al., 1998b, {\sl Astrophys. J.} {\bf 500}, pp. L141-L144.
\bibitem[]{} Stift, M.J., 1995, {\sl Astr. Astrophys. } {\bf 301}, 776-780.
\bibitem[]{} Stothers N.R., 1988, {\sl Astrophys. J.} {\bf 329}, 712-719.
\bibitem[]{} Tanvir N.R. et al., 1995, {\sl Nature  }{\bf 377}, 27-31. 
\bibitem[]{} Tomaney A.B., 1997,  in {\sl Astrophysical returns of microlensing surveys}, ed R. Ferlet, J.P. Maillard, \'editions fronti\`eres, pp. 59-68.
\bibitem[]{} Udalski A. et al., 1992 {\sl Acta Astron.} {\bf 42}, pp. 253-283.
\bibitem[]{} Udalski A. et al., 1994 {\sl Acta Astron.} {\bf 44} pp. 317-386. 
\bibitem[]{} Udalski A. et al., 1995a {\sl Acta Astron.} {\bf 45} pp. 1-236.
\bibitem[]{} Udalski A. et al., 1995b {\sl Acta Astron.} {\bf 45} pp. 433 -622.
\bibitem[]{} Udalski A. et al., 1997 {\sl Acta Astron.} {\bf 47}, pp. 319-344.
\bibitem[]{} Udalski A., 1998a {\sl Acta Astron.} in press
\bibitem[]{} Udalski A. et al., 1998b {\sl Acta Astron.} submitted  
\bibitem[]{} Van Albada T.S. \& Baker N., 1971, {\sl Astrophys. J.} {\bf 169}, pp. 311-322.
\bibitem[]{} Walker, A.R. \& Nemec, J.M., 1996, {\sl Astron. J.} {\bf 112}, pp. 2026-2052.  
\bibitem[]{} Welch D.L. et al., 1997,  in {\sl Astrophysical returns of microlensing surveys},  ed R. Ferlet, J.P. Maillard, \'editions fronti\`eres, pp. 205-212.
\bibitem[]{} Zaritsky, D., et al., 1994, {\sl Astrophys. J.} {\bf 420}, pp. 87-109.
\end{thebibliography}
}
\end{document}